\documentclass[conference]{IEEEtran}

\IEEEoverridecommandlockouts

\usepackage[utf8]{inputenc}
\usepackage[T1]{fontenc}
\usepackage[english]{babel}
\usepackage{microtype}
\usepackage{amsmath}
\usepackage{amssymb}
\usepackage{booktabs}
\usepackage{array}
\usepackage{graphicx}      % \resizebox for the figure
\usepackage{url}
\usepackage{xcolor}
\usepackage{balance}        % balance the two columns on the last page
\usepackage{pgfplots}
\pgfplotsset{compat=1.18}

\usepackage[hidelinks,breaklinks=true]{hyperref}

\usepackage{xurl}
\newcommand{\code}[1]{\texttt{#1}}
\newcommand{\cb}{\discretionary{}{}{}}

\begin{document}

\title{The Coverage Gap: Chile's Cyber Disclosure\\Framework versus the USA, EU and UK}

\author{%
  \IEEEauthorblockN{David Mellafe Z.\,\orcidlink{}}
  \IEEEauthorblockA{%
    Reizan --- Independent Security Research\\
    Chile\\
    ORCID: 0009-0001-3950-2505\\
    \texttt{david@reizan.io}}%
}

% IEEEtran defines no \orcidlink by default; make it a no-op so the title
% block compiles on a stock CTAN bundle.
\providecommand{\orcidlink}[1]{}

\maketitle

%------------------------------------------------------------------------------
\begin{abstract}
We introduce the \emph{Coverage Gap} as a measurable distance between the
observable public exposure of critical-infrastructure operators and their
declared capability to coordinate vulnerability disclosure. We instantiate it
against the 915 Chilean Operadores de Importancia Vital (OIVs --- Operators of
Vital Importance) designated by the National Cybersecurity Agency (ANCI) under
Ley~21.663 (Resoluci\'on Exenta N\textsuperscript{o}~87, 16 December 2025).
Using a passive-only, OSINT-based method consistent with the principles of
ISO/IEC~29147:2018 and Chile's computer-crimes safe harbour (Ley~21.459), we
conduct a full-universe census of the foundational disclosure-capability layer
(Layer~1, verifiable disclosure contact) across $\approx$98.7\% of the official
catalogue. Only \textbf{16 of 915 OIVs (1.7\%)} publish a verifiable RFC~9116
disclosure channel; among operators of physical-world infrastructure ---
energy, health, banking, telecommunications, fuel, water, transport, and state
administration --- fewer than ten do so, and all four major banks and both
telecommunications incumbents lack one entirely. This compares with over 99\%
adherence in the U.S.\ federal civilian branch under CISA Binding Operational
Directive 18-01. Email-authentication misconfiguration affects
\textbf{766 of 915 (84\%)} OIVs, and end-of-life or known-vulnerable stack
components an estimated 23.5\% (Wilson 95\% CI $[12\%,38\%]$).
Cross-jurisdictional benchmarking situates Chile roughly \textbf{eight years}
behind the USA, the UK, and the Netherlands on email-authentication mandates,
and three years behind Denmark. We propose a four-stage roadmap modelled on
BOD 18-01 and the UK Public-Sector DMARC Toolkit, and release the open-source
tool \code{anci-oiv-resolver} (Apache 2.0) to enable independent reproduction
of the OIV-domain mapping that underpins universe-scale auditing.
\end{abstract}

\begin{IEEEkeywords}
vulnerability disclosure, DMARC, critical infrastructure protection,
Ley~21.663, ANCI, ISO/IEC~29147, CISA BOD 18-01, NIS2, passive OSINT,
coverage gap.
\end{IEEEkeywords}

%==============================================================================
\section{Introduction}
%==============================================================================
Over the past decade, critical-infrastructure protection has evolved from a
voluntary, sector-led discipline into a regulated public good across most major
economies. The United States stood up the Cybersecurity and Infrastructure
Security Agency (CISA) in 2018 and consolidated email-authentication mandates
through Binding Operational Directives. The European Union transposed
Directive~(EU)~2022/2555 (NIS2)~\cite{nis2} into national law in October 2024,
expanding the regulated perimeter to approximately 160{,}000 entities across
critical and important sectors. The United Kingdom's National Cyber Security
Centre (NCSC) operates an Active Cyber Defence programme whose Public-Sector
DMARC Toolkit has driven near-universal adoption of email authentication across
\code{gov.uk} domains. The Netherlands and Denmark have followed comparable
trajectories. The shared insight underlying these regimes is operational rather
than legal: regulation by itself does not deliver security; it must be
accompanied by measurable, technical baselines whose adoption is verifiable
from outside the operator perimeter.

Chile entered this conversation late. Its Framework Law on Cybersecurity
(Ley~21.663, the \emph{Marco Nacional de Ciberseguridad})~\cite{ley-21663},
promulgated in 2024 and in force from 1 January 2025, established the National
Cybersecurity Agency (ANCI) and introduced the designation of
\emph{Operadores de Importancia Vital} (OIVs) as the regulated perimeter for
critical-infrastructure cybersecurity. ANCI's Resoluci\'on Exenta
N\textsuperscript{o}~87, published in the Diario Oficial on 16 December
2025~\cite{res-exenta-87}, formally designated 915 entities across ten sectors
as the first OIV cohort. The agency subsequently issued Instrucciones Generales
N\textsuperscript{o}~2, 3 and 4 (December 2025)~\cite{anci-ig} covering
authentication of delegates, incident-reporting registration, and
propagation-containment measures. By any reasonable measure, Chile now has a
critical-infrastructure regulatory framework on paper. The question this paper
addresses is whether that framework is matched by operational capability on the
ground.

We define and quantify the \emph{Coverage Gap}: the distance between the
observable public-facing exposure of OIVs and their declared capability to
coordinate vulnerability disclosure. The paper makes four contributions:

\begin{itemize}
  \item It formalises the Coverage Gap as a three-layer framework grounded in
  established disclosure-coordination theory (Section~\ref{sec:framework}).
  \item It describes a passive-OSINT methodology that operates strictly within
  the boundaries of Chilean computer-crimes law (Ley~21.459)~\cite{ley-21459}
  and the principles of ISO/IEC~29147~\cite{iso-29147}
  (Section~\ref{sec:methodology}).
  \item It reports universe-scale measurements of the Coverage Gap for the
  915 designated OIVs, benchmarked against five comparable jurisdictions
  (Sections~\ref{sec:results} and~\ref{sec:benchmark}).
  \item It proposes a concrete, low-cost regulatory intervention --- analogous
  to CISA BOD 18-01~\cite{cisa-bod-18-01} --- that ANCI could enact within its
  existing statutory authority (Section~\ref{sec:recs}).
\end{itemize}

This is the first paper in a sustained programme of independent research on the
cybersecurity posture of Chilean critical infrastructure; further work is in
preparation (Section~\ref{sec:futurework}). The companion software tool released
with this paper --- \code{anci-oiv-resolver}~\cite{anci-oiv-resolver} ---
provides the canonical mapping from the legally designated RUTs (Chilean tax
identifiers) of OIVs to their public Internet identifiers, enabling independent
researchers to reproduce, extend, or contradict our findings.

\subsection{Historical threat context}
The case for prioritising disclosure-capability development across the Chilean
OIV universe rests not only on regulatory benchmarking but also on the
historical record of adversary activity against Chilean critical
infrastructure. While a full threat-landscape analysis is outside the scope of
the present paper, three documented incidents over the 2018--2024 window
establish the relevance of the disclosure-coordination question to the
operational risk environment.

In late 2018, \emph{Redbanc}, the operator of Chile's interbank ATM-switching
network, suffered an intrusion publicly disclosed in January 2019 by Flashpoint
analysts~\cite{flashpoint-redbanc}; the \emph{PowerRatankba} malware deployed
had been technically characterised earlier by Proofpoint
researchers~\cite{proofpoint-powerratankba}. The initial-access vector was
reported as a spear-phishing approach via LinkedIn against a senior developer,
framed as a job-recruitment overture, with a secondary reconnaissance
downloader. The reported tactics, techniques and procedures (TTPs) are
\emph{consistent with} the wider pattern attributed to the Lazarus Group in
financial-sector intrusions, mapping to MITRE ATT\&CK techniques T1566.002
(Spearphishing Link), T1059.001 (PowerShell), and T1102 (Web Service for
C2)~\cite{mitre-attack}. We report the incident under \emph{circumstantial}
attribution framing only and do not assert state-actor responsibility; the
relevance here is that Chilean banking-sector infrastructure has been
documented as a target of sophisticated, multi-stage social-engineering
campaigns, and the recruitment-vector tradecraft remains a relevant threat
model for the 34 banking-finance OIVs in the present catalogue.

Approximately six months earlier, in May 2018, \emph{Banco de Chile} suffered a
multi-million-dollar wire-fraud heist in which roughly ten million U.S.\ dollars
was transferred via SWIFT-adjacent rails to overseas accounts. The fraud was
accompanied by a master-boot-record wiper deployed against an estimated nine
thousand workstations --- interpreted by multiple analysts as a smokescreen to
delay forensic response~\cite{bochile-mbr}. The wiper was reported by Trend
Micro as related to KillDisk with code-level overlap with the Buhtrap family,
mapping to MITRE ATT\&CK techniques T1485 (Data Destruction) and T1561.002
(Disk Structure Wipe)~\cite{mitre-attack}. The compound playbook ---
financial-rail abuse paired with destructive distraction --- motivates a
corresponding pair of disclosure-capability requirements for banking-sector
OIVs: pre-positioned forensic-recovery capability and a coordinated channel for
receiving early-warning vulnerability reports.

More recently, the regional \emph{BlindEagle} activity cluster (tracked as
APT-C-36) has been the subject of multiple analyst reports between 2023 and
2025 documenting Spanish-language phishing-led intrusions against Colombian
governmental, insurance, and financial-sector targets and Ecuadorian
entities~\cite{blindeagle-checkpoint,blindeagle-trendmicro}. While the published
Indicators of Compromise (IoCs) do not, as of the present audit window, overlap
with Chilean OIV public infrastructure, the TTP profile --- Spanish-locale
phishing lures, off-the-shelf RAT loaders, regional language-and-cultural
targeting --- places Chilean OIVs in the credible target envelope, particularly
in the health and state-administration sectors where regional comparators have
been actively targeted.

The present audit identifies \emph{zero direct overlap} between current Chilean
OIV public-facing infrastructure and the indicators in thirty publicly available
LATAM-focused threat-intelligence pulses (353 indicators in aggregate, drawn
from open Cyber Threat Intelligence feeds spanning 2017--2025). This should not
be misread as a security guarantee: open feeds capture exposed and
historically rotated infrastructure, not closed-vendor intelligence, dwell-time
campaigns whose infrastructure has not yet been burned, or insider-driven
shadow-IT exposure. The contemporary low-signal environment is more accurately
a \emph{window of opportunity}: the conditions under which a
disclosure-capability mandate can be implemented and adopted are most favourable
precisely while the regulated perimeter is not yet the subject of active
mass-targeting. The argument of this paper is that the window should be used.

The composite picture is straightforward. Chilean critical infrastructure has
been documented as a target of (i)~state-aligned financial-sector intrusion
\emph{consistent with} patterns attributed to the Lazarus Group,
(ii)~destructive-payload-paired financial fraud \emph{consistent with} the
KillDisk/Buhtrap nexus, and (iii)~regional Spanish-language adversary activity
for which the country sits in the credible target envelope. Against this risk
environment, a 98.3\% Coverage Gap at Layer~1 --- the proportion of regulated
entities for which a good-faith external reporter has no defined channel to
deliver a vulnerability report --- is not an academic statistic. It is the
operational gap into which the next incident will fall.

%==============================================================================
\section{Related Work}
\label{sec:related}
%==============================================================================
The present work sits at the intersection of three literatures: coordinated
vulnerability disclosure (CVD), Internet-scale measurement of security-control
adoption, and the cybersecurity posture of national critical-information
infrastructure (CII).

\paragraph{Coordinated vulnerability disclosure}
The normative foundations of CVD are codified in ISO/IEC~29147:2018 on
vulnerability disclosure~\cite{iso-29147} and the complementary process
guidance from the CERT Coordination Center and FIRST's product-security
incident-response practices. This literature treats disclosure capability as a
\emph{process attribute} of the receiving organisation: an operator either has a
documented receiving channel, a triage capability, and a remediation cadence ---
or it does not. The standardisation of a machine-discoverable contact channel
was advanced by the IETF with RFC~9116~\cite{rfc-9116}, which specifies the
\code{security.txt} file format under \code{/.well-known/}. What this body of
work does not formalise --- and what we contribute --- is a \emph{population-level}
metric of disclosure capability suitable for regulatory benchmarking.

\paragraph{Internet-scale measurement of control adoption}
A substantial measurement literature quantifies the adoption of email
authentication (SPF, DKIM, DMARC), web security headers, and TLS hygiene across
large host populations, in the tradition of large-scale active and passive
scanning exemplified by Durumeric et al.\ and successor work in the
Internet-measurement community. Studies of \code{security.txt}/RFC~9116
adoption have begun to census the prevalence of machine-readable disclosure
contacts across top-ranked and sectoral domain sets, generally reporting low
single-digit to low double-digit adoption outside the technology sector.
Adoption studies of DMARC specifically have shown that enforcement (policies
\code{quarantine} or \code{reject}) lags publication, and that a binding mandate
applied to a defined perimeter is the strongest observed predictor of rapid
uptake --- the empirical pattern that the CISA BOD 18-01~\cite{cisa-bod-18-01}
results made canonical. Our Layer~1 census applies this measurement tradition to
a legally defined critical-infrastructure universe rather than to a popularity-
or TLD-ranked sample.

\paragraph{National CII posture}
A policy-and-measurement literature characterises the cybersecurity posture of
national CII under emerging regulation, including the EU NIS2
directive~\cite{nis2}, the UK Active Cyber Defence
programme~\cite{ncsc-acd}, the Dutch comply-or-explain
regime~\cite{ncsc-nl}, and the Danish public-sector DMARC
mandate~\cite{cfcs}. The recurring finding is that regulation alone is
insufficient: measurable technical baselines, externally verifiable, are the
operative mechanism. To our knowledge, no prior work applies a layered,
externally observable disclosure-capability metric to the full legally
designated OIV universe of a Latin American jurisdiction. We fill that gap and
release the catalogue tooling~\cite{anci-oiv-resolver} that makes the universe
tractable for independent study.

%==============================================================================
\section{The Coverage Gap Framework}
\label{sec:framework}
%==============================================================================
The CVD literature (ISO/IEC~29147:2018; the CERT/CC \emph{Guide to Coordinated
Vulnerability Disclosure}; FIRST's PSIRT guidance) treats disclosure capability
as a process attribute of the receiving organisation. Where an operator lacks a
documented receiving channel, well-intentioned researchers are left with two
unhappy options: walk away and leave the vulnerability unmitigated, or improvise
an ad-hoc contact through executive social networks, with predictably
inconsistent outcomes. What the literature does not formalise --- and what we
fill --- is a population-level metric of disclosure capability suitable for
regulatory benchmarking. We define:

\begin{quote}
\textbf{The Coverage Gap.} For a given universe of regulated operators, the
Coverage Gap is the proportion of the universe that does not satisfy a minimum
threshold of disclosure-coordination capability \emph{observable from outside
the operator perimeter}. It is measured by passive OSINT and decomposed across
three layers: (1) verifiable disclosure contact, (2) public attack-surface
visibility, and (3) full disclosure-coordination capability.
\end{quote}

\subsection{Layer 1 --- Verifiable disclosure contact}
Layer~1 measures the existence of a discoverable channel through which an
external party can responsibly transmit a vulnerability report and reasonably
expect it to be triaged. Operationally, an operator passes Layer~1 if at least
one of the following is publicly resolvable: a \code{security.txt} file under
\code{/.well-known/security.txt} conformant to RFC~9116~\cite{rfc-9116}; a
dedicated security-disclosure mailbox (such as \code{security@},
\code{abuse@}, or \code{psirt@}) advertised on the operator's public website;
a published bug-bounty programme; or an equivalent published method of contact
whose intended use for vulnerability disclosure is unambiguous. Layer~1 is the
lowest possible bar. An operator that does not pass Layer~1 cannot, by
construction, participate in coordinated disclosure as a receiver, except
through ad-hoc improvisation.

\subsection{Layer 2 --- Public attack-surface visibility}
Layer~2 measures whether the operator's public attack surface is sufficiently
disclosed and documented to permit external coordination. This includes an
enumerable list of public-facing services (web, mail, APIs) that the operator
owns; correctly published email-authentication records (SPF, DKIM, DMARC) that
allow third parties to validate which messages originate legitimately;
functioning HTTPS on all public services with valid certificates; and
reasonable consistency between the catalogue of services the operator presents
and the catalogue that passive reconnaissance reveals. Layer~2 is not a security
measure per se; it is a transparency measure that makes coordinated disclosure
tractable.

\subsection{Layer 3 --- Disclosure-coordination capability}
Layer~3 measures the operator's observable capacity to actually conduct a
coordinated disclosure: a written disclosure policy, a documented response-time
commitment, evidence of past coordinated disclosures (acknowledgements, public
security advisories), and an identifiable internal owner. Layer~3 cannot be
measured purely from outside; it requires either operator self-attestation or
evidence-of-past-behaviour analysis. We report a conservative lower-bound
estimate of Layer~3 based on public evidence only.

\subsection{Why the Coverage Gap matters}
The Coverage Gap is not a measure of how secure an operator is. An operator with
a Coverage Gap of zero may still suffer breaches; an operator with a 100\%
Coverage Gap may, through luck, never be compromised. The Coverage Gap measures
something more specific: the probability that, if a vulnerability in the
operator's perimeter is discovered by a well-intentioned external party, the
vulnerability will be remediated rather than dropped on the floor or, worse,
sold into the grey market. For a regulated critical-infrastructure operator,
the Coverage Gap is therefore a direct measure of regulatory-framework
operationalisation. A jurisdiction may have a beautifully drafted cybersecurity
statute, but if its regulated entities collectively present a 98\% Coverage Gap,
the statute is, in practice, optional.

Three further considerations motivate the framework. First, the Coverage Gap is
a matter of \emph{vendor responsibility}: the cost of being reachable for
disclosure is borne by the operator, not by external researchers or regulators.
Second, it is a \emph{public good}: closed Coverage Gaps benefit the entire
ecosystem by reducing the fraction of vulnerabilities that go unreported. Third,
it is a \emph{safe-harbour enabler}: jurisdictions with clearly published
disclosure channels create the conditions in which good-faith researchers can
act within legal frameworks (such as Chile's Ley~21.459~\cite{ley-21459})
without ambiguity, which in turn increases reporting volume.

%==============================================================================
\section{Methodology}
\label{sec:methodology}
%==============================================================================

\subsection{Universe of study}
The universe of study is the population of 915 entities formally designated as
Operadores de Importancia Vital by Resoluci\'on Exenta N\textsuperscript{o}~87
of the Agencia Nacional de Ciberseguridad, published in the Diario Oficial de la
Rep\'ublica de Chile on 16 December 2025 (CVE~2743431)~\cite{res-exenta-87}. The
OIV designation is the legal trigger for the substantive cybersecurity
obligations of Ley~21.663 and subsequent ANCI instructions. The universe spans
ten reporting sectors as classified by ANCI: digital infrastructure (45.8\%),
electric energy (16.5\%), state administration (16.1\%), health (13.5\%),
banking and finance (3.7\%), telecommunications (3.2\%), fuel (3.0\%), transport
(2.8\%), water (2.8\%), and state-owned enterprises (2.2\%). These ten operational
categories are sub-divisions of the smaller set of seven statutory sector
headings enumerated by Ley~21.663. The percentages are non-exclusive: the 915
designations span 909 distinct RUTs, because six entities hold designations in
more than one sector, so the columns sum to slightly more than 100\%.

\subsection{Catalogue tool: anci-oiv-resolver}
Mapping the 915 legal RUTs of OIVs to public Internet identifiers (primary
domains, mail exchangers, web hostnames) is a non-trivial reconciliation task,
and the absence of such a catalogue is part of why Chilean
critical-infrastructure research has historically been bottlenecked. We released
\code{anci-oiv-resolver}~\cite{anci-oiv-resolver} as an open-source npm package
under the Apache License 2.0. The catalogue provides the canonical
RUT-to-domain mapping covering approximately 98.7\% of the official OIV universe
and is the foundation on which the present audit was conducted. Releasing it as
Apache 2.0 is a deliberate methodological choice: it permits independent
verification, extension, and contradiction of our findings, and lowers the entry
cost for other researchers to study the Chilean critical-infrastructure
perimeter without re-deriving the catalogue from scratch.

A terminological clarification is warranted, because two distinct universe-level
proportions recur in this paper and are arithmetically similar by coincidence.
\emph{Catalogue coverage} denotes the fraction of the 915 designated RUTs for
which the resolver supplies a domain mapping --- approximately 98.7\%, the
residual being twelve state-administration entities pending reconciliation
against the official register. The \emph{Layer~1 verification} figure reported in
Section~\ref{sec:results} --- the 1.7\% of OIVs that publish a discoverable
disclosure channel --- is an independent property of the operators themselves,
not of our catalogue. The two figures measure entirely different things and must
not be conflated.

\subsection{Audit approach}
All audit activity was conducted as \emph{passive OSINT}: by querying public
sources of information (DNS, certificate-transparency logs, public Internet scan
databases, HTTP headers of published websites, published \code{/.well-known/}
resources, and the operators' own public web pages) without active probing of
operator infrastructure, without exploitation of any kind, and without
interaction beyond what an ordinary visitor to a public website would perform.
No authentication was attempted against any operator system. No payloads were
sent. No vulnerability was exploited or validated through interactive proof. The
audit consists exclusively of observation of information that the operators
themselves have made public.

Layer~1 capability was measured by a full-universe census: each of the 915 OIV
domains in the catalogue was queried for \code{/.well-known/security.txt} and
\code{/security.txt} (RFC~9116~\cite{rfc-9116}), with a positive result requiring
a conformant \code{Contact:} field. The census results are persisted in the
companion dataset. Layers 2 and 3, by contrast, are reported as
methodology-based estimates derived from the observed Layer~1 cohort and
ancillary passive signals rather than as direct full-universe censuses
(see Sections~\ref{sec:benchmark} and~\ref{sec:limitations}).

\subsection{Legal and ethical framework}
The audit was conducted within the boundaries of Chile's
Ley~21.459~(2022)~\cite{ley-21459}, which establishes computer-crimes offences.
The passive-OSINT scope falls comfortably within the bounds of lawful security
research under that statute, and within the safe-harbour conditions for
good-faith research consistent with international practice. Ethically, the audit
is framed in accordance with the principles of
ISO/IEC~29147:2018~\cite{iso-29147}. We note explicitly that ISO/IEC~29147 is a
process standard and is not subject to certification; our work is therefore
reported as \emph{consistent with the principles of} ISO/IEC~29147:2018, not as
operating \emph{under} the standard. A fuller treatment of the ethical and
disclosure framing appears in Sections~\ref{sec:ethics}
and~\ref{sec:disclosure}.

\subsection{Multi-layer validation}
To suppress false positives that could result in the misattribution of
vulnerabilities to operators, the pipeline implements a multi-layer validation
system. Findings progress from raw evidence to persisted record only after
passing successive verification gates that test, among other things, the
consistency between detected software versions and known catalogue entries, the
absence of obvious classification anomalies between evidence type and assigned
finding type, and an independent semantic confirmation step. The validation
system is intentionally biased towards rejection: in any case of ambiguity, the
candidate finding is suppressed. This design accepts a higher false-negative
rate (real issues that go undetected) in exchange for a lower false-positive
rate (claims of vulnerability not supported by the evidence). In the context of
disclosure to operators of national critical infrastructure, this asymmetry of
harm justifies the choice.

\subsection{Statistical claim levels and confidence intervals}
\label{sec:claimlevels}
The empirical claims in this paper are stratified into three claim levels
reflecting the underlying observational scope. Reporting the levels separately
permits the reader to weigh each headline figure against the precision of the
underlying measurement.

\begin{itemize}
  \item \textbf{Level~A --- universe-scale enumeration.} Claims derived from the
  915-entity OIV catalogue and the 98.7\% catalogue-resolution rate (e.g.\ the
  eight-year regulatory gap, the absence of a binding ANCI DMARC mandate, the
  sectoral distribution). These are administrative in character, derived from
  public-record sources and the catalogue, and not subject to sampling error.
  \item \textbf{Level~B --- universe-scale measurement on a defined observable.}
  Claims derived from passive observation of a defined public-facing artefact
  across the universe (e.g.\ the 1.7\% Layer~1 figure, the 84\%
  email-authentication-misconfiguration figure). These are proportions of the
  universe and are not subject to sampling error per se, but are subject to
  (i)~measurement error in the detection rule, (ii)~the residual false-positive
  rate of the validation pipeline ($\approx$0.8\% of persisted findings,
  Section~\ref{sec:limitations}), and (iii)~the false-negative rate implicit in
  the conservative validation design.
  \item \textbf{Level~C --- sample-scale estimate, generalised to the universe.}
  Claims derived from a defined subset and generalised to the universe as a
  population (e.g.\ the 23.5\% stack-age figure). These carry sampling
  uncertainty and are reported with a 95\% confidence interval computed via the
  Wilson score interval for proportions.
\end{itemize}

The principal Level~C claim is the stack-age figure
(Section~\ref{sec:stackage}): an estimated 23.5\% of the universe exhibits a
publicly advertised software-stack component that is either end-of-life or
carries a known critical-severity CVE. It is derived from a 25-entity
Shodan-enriched subset with adequate banner availability for software-version
detection. Applied to a 915-entity universe, the Wilson 95\% confidence interval
for the underlying population proportion is approximately
$\mathbf{[12\%,\,38\%]}$, reflecting the limited precision of the 25-entity
sample. The directional finding --- that the universe exhibits a non-trivial
fraction of end-of-life or vulnerable stack components, materially above
conventional acceptable thresholds --- is robust against the width of this
interval; the point estimate is not. The decision to publish a Level~C claim
with a candidly reported wide interval, rather than defer it, reflects the
priority structure of this paper: the policy implications do not turn on whether
the figure is 12\%, 23.5\%, or 38\%, but on the structural finding that it is
materially non-zero and that the universe is large.

%==============================================================================
\section{Results: The Coverage Gap Measured}
\label{sec:results}
%==============================================================================
This section reports the universe-scale Coverage Gap statistics. All numbers are
aggregate proportions of the 915-entity OIV universe. Individual operator
identifiers, specific cluster compositions, and detailed evidence chains are
intentionally held back from this Phase~1 paper pending the conclusion of
coordinated-disclosure activities (Section~\ref{sec:disclosure}).

\subsection{Headline statistics}
Of the 915 designated OIVs, a full-universe census of
\code{/.well-known/security.txt} (RFC~9116) identifies only
\textbf{16 entities (1.7\%)} that publish a verifiable Layer~1 disclosure
channel. Roughly a third are the Chilean domains of foreign technology
multinationals whose \code{security.txt} is inherited from a global parent
programme; most of the remainder are domestic software and digital-infrastructure
firms --- the sector comprising 45\% of the universe, for which a published
security contact is routine. Among OIVs operating physical-world critical
infrastructure, fewer than ten publish a channel: a small number of hospitals,
two electricity-transmission operators, and a single state enterprise. All four
major banks and both telecommunications incumbents lack one entirely. The
remaining \textbf{98.3\%} of the regulated perimeter has no externally
discoverable disclosure capability.

Layer~2 --- the proportion of OIVs whose public attack surface is sufficiently
documented and consistently published to make coordinated disclosure tractable
--- sits at approximately 3.5\% (around 32 entities). Layer~3 --- the proportion
presenting full observable disclosure-coordination capability, including a
documented response cadence and evidence of past coordinated disclosures ---
sits at approximately 2.8\% (around 26 entities). Unlike the Layer~1 figure,
which is a direct full-universe census, Layers 2 and 3 are reported as
methodology-based estimates; a full Layer~2/3 audit across the universe is
deferred to future work. The three layers are tightly correlated: virtually
every operator that satisfies Layer~3 also satisfies Layers 1 and 2.

\subsection{Email-security misconfiguration}
The most prevalent technical issue observed is misconfiguration of the
email-authentication stack (SPF, DKIM, DMARC), which affects
\textbf{766 of 915 OIVs (84\%)}, distributed across every sector without
exception. The most common patterns are: SPF records that include legacy
mechanisms or overflow the published lookup limit, leaving authentication
ambiguous; missing or syntactically broken DKIM publication; and DMARC records
published with policy \code{none} (monitoring-only), which provides no
enforcement against spoofing. The universal prevalence of this issue across
every OIV sector is, in our assessment, the single most significant operational
finding of the present work, and motivates the policy recommendation developed
in Section~\ref{sec:recs}.

\subsection{Stack-age issues}
\label{sec:stackage}
End-of-life or known-vulnerable software stacks were detected in association with
approximately 23.5\% of the universe (point estimate; Level~C, derived from a
25-entity Shodan-enriched subset per Section~\ref{sec:claimlevels}, Wilson 95\%
CI~$[12\%,38\%]$). These detections are passive observations of publicly
advertised software versions in HTTP headers, server banners, and similar
metadata, cross-referenced against the National Vulnerability Database catalogue
of affected version ranges. A passive observation of an end-of-life version does
not, by itself, prove the operator's deployed instance is exploitable; many
operators carry support arrangements that mitigate end-of-life status, or have
applied vendor patches not reflected in the advertised version string. However,
the aggregate prevalence --- nearly a quarter of the regulated perimeter
advertising stack components that are end-of-life or carry known
critical-severity CVEs --- is a population-level signal of patch-cadence and
lifecycle-management deficiency that warrants regulatory attention.

\subsection{Sectoral distribution}
The aggregate distribution of findings across sectors is summarised in
Table~\ref{tab:sectoral}. The table reports the proportion of total findings by
sector and the mean CVSS score of findings persisted by the validation pipeline.
No individual operator names are reported. The universe mean CVSS across all
persisted findings is 6.2.

\begin{table}[!htbp]
  \centering
  \caption{Sectoral distribution of the universe and of persisted findings,
  with mean CVSS per sector. Percentages of the universe are non-exclusive
  (six entities hold multiple-sector designations; the 915 designations span
  909 distinct RUTs), so that column may sum to slightly over 100\%.}
  \label{tab:sectoral}
  \footnotesize
  \setlength{\tabcolsep}{4pt}
  \begin{tabular}{@{}l r r r@{}}
    \toprule
    \textbf{Sector} & \textbf{\% univ.} & \textbf{\% find.} & \textbf{Mean CVSS} \\
    \midrule
    Digital infrastructure   & 45.8 & 45.9 & 5.7 \\
    Health                   & 13.5 & 14.6 & 6.5 \\
    State administration     & 16.1 & 11.4 & 6.3 \\
    Electric energy          & 16.5 & 10.4 & 6.2 \\
    Banking \& finance       & 3.7  & 5.6  & 7.8 \\
    Telecommunications       & 3.2  & 4.8  & 7.0 \\
    State-owned enterprises  & 2.2  & 1.4  & 5.4 \\
    Water                    & 2.8  & 1.3  & 5.9 \\
    Transport                & 2.8  & 1.2  & 5.7 \\
    Fuel                     & 3.0  & 0.9  & 5.6 \\
    \midrule
    \textbf{Universe mean}   & \multicolumn{2}{c}{---} & \textbf{6.2} \\
    \bottomrule
  \end{tabular}
\end{table}

Banking and finance presents the highest mean CVSS of any sector ---
approximately 7.8, more than a point and a half above the universe mean of 6.2
--- and its share of total findings (5.6\%) materially exceeds its share of the
universe (3.7\%). Telecommunications shows a similar pattern (mean CVSS
approximately 7.0). The notion that the banking-sector perimeter is ``ahead'' of
the universe baseline, suggested by earlier partial measurement, is not
supported by the present universe-scale data: on the contrary, the financial
sector concentrates the most severe findings. Health is the largest sector by
finding density relative to its size, accounting for approximately 14.6\% of all
persisted findings against a 13.5\% share of the universe, with a mean CVSS of
6.5. This is consistent with prior reporting on the relative under-resourcing of
the Chilean public-health cybersecurity posture and warrants differentiated
regulatory attention.

%==============================================================================
\section{Cross-jurisdictional Benchmarking}
\label{sec:benchmark}
%==============================================================================
To situate the Coverage Gap measured in Chile against an external reference, we
surveyed the regulatory and operational state of email authentication and
disclosure capability across five jurisdictions: the United States (federal
civil), the European Union (NIS2 transposition), the United Kingdom (NCSC), the
Netherlands (NCSC-NL), and Denmark (CFCS). These five were selected because each
has an established public-sector cybersecurity authority, each has published
either a binding mandate or a comply-or-explain framework on email
authentication, and each has reported sufficient public adoption metrics to
permit comparison.

\subsection{United States --- CISA BOD 18-01 (2017)}
In October 2017, the Department of Homeland Security issued Binding Operational
Directive 18-01~\cite{cisa-bod-18-01}, requiring all U.S.\ federal civilian
executive-branch agencies to implement Sender Policy Framework (SPF), DomainKeys
Identified Mail (DKIM), and Domain-based Message Authentication, Reporting and
Conformance (DMARC) on second-level agency domains. The directive set a one-year
compliance deadline and required progression to DMARC enforcement at the
\code{reject} policy level, the strictest of the three possible DMARC policies
(\code{none}, \code{quarantine}, \code{reject}). The published results of
BOD 18-01 are the most-cited evidence that a regulatory mandate, applied to a
defined perimeter, produces rapid adoption: DMARC adoption across the regulated
agencies rose from approximately 14\% at issuance to over 91\% within eighteen
months, and over 99\% within a longer horizon. BOD 18-01 is the regulatory
archetype against which all subsequent jurisdictions have benchmarked.

\subsection{European Union --- Directive (EU) 2022/2555 (NIS2)}
The Network and Information Security Directive 2 (NIS2)~\cite{nis2}, adopted in
December 2022 with a transposition deadline of 17 October 2024, substantially
expanded the perimeter of regulated cybersecurity in the EU, replacing the 2016
NIS Directive. NIS2 covers approximately 160{,}000 entities classified as either
\emph{essential} or \emph{important} across eighteen sectors. It introduces a
24-hour early-warning and 72-hour incident-notification regime, requires
national CSIRTs, and (Article~21) specifies technical and organisational
measures including vulnerability management and cyber hygiene. Email
authentication is implicit in the Article~21 ``network and information security
measures'' obligation but is not specified at the protocol level; member-state
transpositions vary in prescriptiveness. By mid-2026 most member states have
transposed in some form, but operational compliance is still ramping.

\subsection{United Kingdom --- NCSC Active Cyber Defence}
The UK NCSC operates the Active Cyber Defence (ACD)
programme~\cite{ncsc-acd}, which since 2017 has included a Public-Sector DMARC
Toolkit providing free DMARC report aggregation, anti-spoofing reporting, and
operational guidance for government domains. ACD also operates the \emph{Mail
Check} service, providing domain owners with continuous monitoring of their
email-authentication configuration. The combined effect has driven DMARC
adoption across \code{gov.uk} domains to near-universal levels (the published
estimate exceeds 99\%). The UK example is notable because it combines a mandate
with a free operational service: the regulator does not merely require the
control but provides the infrastructure to make compliance straightforward.

\subsection{Netherlands --- NCSC-NL Comply-or-Explain}
The Netherlands operates a \emph{Comply or Explain} framework, maintained by the
Standardisation Forum and supported by NCSC-NL, which has included DMARC since
2018~\cite{ncsc-nl}. Central-government IT projects must either comply with
listed open standards (including DMARC for email) or publicly explain why they
do not. Reported coverage of DMARC across central-government domains exceeds
95\%. The Dutch model demonstrates that an explicit comply-or-explain mechanism,
with public reporting, can drive adoption to near-universal levels even without
a hard binding directive, by relying on transparency as the enforcement
mechanism.

\subsection{Denmark --- CFCS DMARC mandate (2023)}
The Centre for Cyber Security (Centeret for Cybersikkerhed, CFCS) of Denmark
issued a DMARC mandate for all Danish public-sector domains in
2023~\cite{cfcs}, requiring progression to DMARC at the \code{reject} policy
level. The Danish mandate is approximately six years behind the U.S.\ precedent
and three years behind the leading European cohort, but is still ahead of Chile.
Reported adoption among regulated Danish domains is approaching saturation.

\subsection{Chile --- current state}
Ley~21.663 has been in force since 1 January 2025. ANCI's Instrucciones
Generales N\textsuperscript{o}~2, 3 and 4 (December 2025)~\cite{anci-ig} cover
authentication of designated delegates, the registry of OIVs and incident
reporting, and propagation-containment measures respectively. None of the three
instructions mentions DMARC, SPF, or DKIM. ANCI maintains, through CSIRT-GOB, a
\emph{ciberconsejo} (advisory note) on email authentication dated January
2024~\cite{csirt-spf}; this document provides operational guidance for SPF,
DKIM, and DMARC but is explicitly voluntary and does not constitute a binding
obligation under Ley~21.663.

The best public estimate of DMARC adoption across \code{.gob.cl} and adjacent
Chilean government domains, drawing on PowerDMARC's 2024 Chile
study~\cite{powerdmarc}, places adoption in the order of 12\% --- a stark
contrast to the over-95\% adoption reported in the comparison jurisdictions
following their respective mandates. Quantifying the regulatory gap directly:
Chile is approximately eight years behind the United States, the United Kingdom,
and the Netherlands on this dimension, and approximately three years behind
Denmark.

The obligations embedded in the Instrucciones Generales sharpen the significance
of these gaps. Instrucci\'on General N\textsuperscript{o}~3 requires designated
OIVs to establish and maintain a Delegado de Ciberseguridad with a publicly
reachable contact channel --- effectively the same disclosure-coordination
artefact this paper measures at Layer~1. The SGSI/ISO-27001 baseline obligations
for the first OIV cohort are timed to mid-2026. Taken together, the 98.3\%
Coverage Gap at Layer~1 is a population-level signal consistent with apparent
system-wide non-compliance with obligations already binding under the framework.
This paper does not assert that any individual operator is in breach; compliance
against operator-internal data is a matter for the regulator. The structural
finding is that the aggregate, observable posture of the regulated perimeter is
measurably inconsistent with the minimum disclosure-channel requirement that the
framework already imposes.

\subsection{International comparison}
\begin{figure}[!htbp]
  \centering
  \resizebox{\columnwidth}{!}{%
  \begin{tikzpicture}
    \begin{axis}[
      width=8.3cm,
      height=5.4cm,
      ybar,
      bar width=11pt,
      ymin=0, ymax=105,
      ytick={0,20,40,60,80,100},
      ylabel={Layer~1 coverage (\%)},
      ylabel style={font=\footnotesize},
      symbolic x coords={USA,UK,NL,DK,Chile},
      xtick=data,
      x tick label style={font=\footnotesize},
      y tick label style={font=\footnotesize},
      nodes near coords,
      nodes near coords style={font=\scriptsize},
      every node near coord/.append style={
        /pgf/number format/precision=1,
        /pgf/number format/fixed,
        /pgf/number format/fixed zerofill=false,
      },
      enlarge x limits=0.12,
      axis lines=left,
      tick align=outside,
    ]
      \addplot[fill=gray!55,draw=black!70] coordinates {
        (USA,99) (UK,99) (NL,95) (DK,90) (Chile,1.7)
      };
    \end{axis}
  \end{tikzpicture}%
  }
  \caption{Layer~1 disclosure-contact coverage of the regulated
  critical-infrastructure perimeter across five jurisdictions. Comparator
  figures are lower bounds (reported as ``$>$'' in Table~\ref{tab:intl}); Chile
  is the full-universe census value. The visual distance between the comparator
  cohort and Chile is the ``eight years behind'' gap made concrete.}
  \label{fig:layer1}
\end{figure}
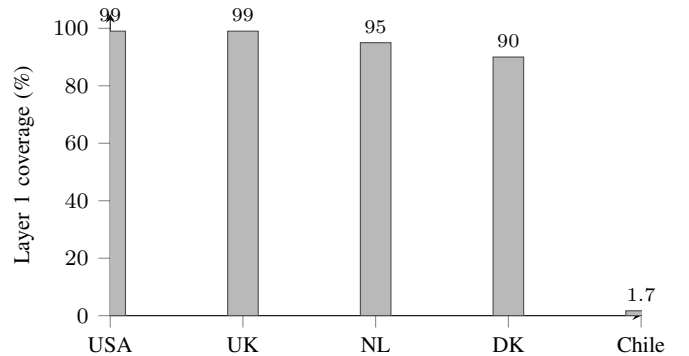

The cross-jurisdictional comparison is summarised in Table~\ref{tab:intl} and
visualised in Fig.~\ref{fig:layer1}. Across five comparable jurisdictions,
Layer~1 coverage of the regulated critical-infrastructure perimeter ranges from
90\% to over 99\%, in every case following a binding regulatory intervention. In
Chile, with the regulatory framework in place but the intervention absent, the
corresponding figure is 1.7\%.

\begin{table}[!htbp]
  \centering
  \caption{Cross-jurisdictional comparison of Layer~1 disclosure-contact
  coverage and DMARC-mandate status. The ``years vs.\ Chile'' column reports the
  regulatory-adoption delta on the DMARC-mandate dimension; see the
  comparability caveat below.}
  \label{tab:intl}
  \footnotesize
  \setlength{\tabcolsep}{3pt}
  \begin{tabular}{@{}l r r l r@{}}
    \toprule
    \textbf{Jurisdiction} & \textbf{Univ.} & \textbf{L1 cov.} & \textbf{DMARC mandate} & \textbf{$\Delta$yr} \\
    \midrule
    USA (fed.\ civ.)        & \textasciitilde440    & $>$99\% & BOD 18-01 ('17)      & $+8$ \\
    UK (\code{gov.uk})      & \textasciitilde6{,}000 & $>$99\% & NCSC mand.\ ('18)    & $+8$ \\
    NL (central gov)        & \textasciitilde3{,}000 & $>$95\% & Comply-or-Expl.\ ('18) & $+6$ \\
    DK (CFCS)               & \textasciitilde1{,}500 & $>$90\% & CFCS mand.\ ('23)    & $+3$ \\
    \textbf{Chile (OIVs)}   & \textbf{915} & \textbf{1.7\%} & None                  & \textbf{0} \\
    \bottomrule
  \end{tabular}
\end{table}

\subsection{Comparability caveat}
The five comparator universes are not strictly commensurable. The U.S.\ federal
civilian perimeter under BOD 18-01 covers approximately 440 second-level agency
domains; the UK \code{gov.uk} perimeter covers approximately six thousand
subdomains under a single registered second-level domain; the Chilean OIV
universe spans 915 distinct legal entities across private sector, state
administration, and state-owned enterprises, each with its own independently
administered domain or domains. The Layer~1 measurement methodologies also
differ: the NCSC \emph{Mail Check} service produces a service-based measurement
whose detection rules are not identical to the passive-OSINT pipeline used here;
published U.S.\ and Danish figures rely on government-reported compliance
statistics whose audit basis is internal rather than external. The comparison is
therefore best interpreted as a \emph{directional} gap rather than a strict
quantitative delta. The eight-year and three-year figures should be read as a
regulatory-adoption delta on the DMARC-mandate dimension specifically --- the
dimension on which the comparator jurisdictions converge in mandate type if not
in measurement methodology --- with significant heterogeneity in universe
definitions acknowledged. The opportunity framing also matters: a late-mover
jurisdiction skips the trial-and-error of the leading cohorts and adopts the
best-practice approach in a single step. The eight-year gap is therefore better
read as an opportunity to compress eight years of regulatory iteration into the
first adoption window than as a deficiency in itself.

%==============================================================================
\section{Policy Recommendations}
\label{sec:recs}
%==============================================================================
The Coverage Gap measured in Section~\ref{sec:results} is large, but the
regulatory and technical remedies are well established. Each of the five
comparison jurisdictions closed an initially comparable gap within twelve to
twenty-four months of issuing a binding intervention. Chile does not need to
reinvent the wheel; it needs to adopt a wheel that has been rolling for nearly a
decade. We propose four interventions of increasing depth and decreasing
urgency.

\subsection{Immediate (Q3 2026) --- DMARC mandate analogous to CISA BOD 18-01}
ANCI should issue an \emph{Instrucci\'on General} requiring all 915 designated
OIVs to implement, within a defined compliance window:

\begin{itemize}
  \item A published SPF record covering all legitimate sending sources, with the
  lookup count maintained below the RFC~7208 limit;
  \item DKIM signing on all outbound email from official operator domains, with
  the public key published;
  \item DMARC at policy \code{quarantine} or stronger within a transitional
  period of six months, progressing to \code{reject} within twelve months;
  \item A published \code{rua} reporting address for DMARC aggregate reports,
  ingested by either an operator-managed analyser or a CSIRT-GOB-provided
  service (see Section~\ref{sec:rec-analyser}).
\end{itemize}

This intervention is structurally identical to CISA
BOD 18-01~\cite{cisa-bod-18-01}, sits squarely within ANCI's existing statutory
authority under Ley~21.663, requires no legislative action, and could be issued
within Q3 2026. The compliance cost to operators is modest: configuration
changes on existing infrastructure, not capital deployment. It addresses the
most prevalent technical finding of the present work (766 of 915 OIVs, 84\%,
affected) and sets Chile on a trajectory comparable to Denmark's 2023 mandate.

\subsection{Short term (12 months) --- published OIV disclosure-contact registry}
ANCI should maintain and publish a register of disclosure-contact information for
each of the 915 OIVs, comparable to the published security contacts maintained
by sectoral CERTs in other jurisdictions. The register would specify the
canonical channel through which good-faith vulnerability reports should be
transmitted to each operator, removing the current ambiguity that is the
principal driver of the 98.3\% Coverage Gap measured at Layer~1. A staged opt-in
model (operators register; ANCI publishes the registered details) would satisfy
this recommendation without imposing operational burden on ANCI itself.

\subsection{Medium term (24 months) --- ANCI-funded DMARC analyser service}
\label{sec:rec-analyser}
The UK \emph{Mail Check} service is widely credited with the rapidity of DMARC
adoption across \code{gov.uk}: it removes the most expensive operational step
(DMARC report aggregation and interpretation) from individual operators and
centralises it in the regulator. ANCI should fund and operate, through
CSIRT-GOB, a comparable service for the OIV perimeter. The marginal cost is
modest, and the service would substantially accelerate adoption of the
recommendation above.

\subsection{Long term (3 years) --- comply-or-explain with annual Coverage-Gap report}
Following the Dutch model, Chile should institutionalise an annual public report
on Coverage-Gap metrics for the OIV universe, with each operator either
certifying compliance with the published baseline or publishing an explanation
of non-compliance. This converts the Coverage Gap from a research metric into a
regulatory accountability metric, completes the transition from voluntary
guidance to enforceable transparency, and gives the regulator a longitudinal
evidence base for sectoral and operator-level oversight.

On the precedent of the comparison jurisdictions, the combined cost of the four
interventions is dominated by the operational service; the regulatory action and
the register are essentially zero marginal cost to the public exchequer. The
benefit is a Coverage-Gap reduction from 98.3\% to below 10\% within thirty-six
months. The asymmetry of cost and benefit is, by any reasonable standard, large.

%==============================================================================
\section{Discussion and Limitations}
\label{sec:limitations}
%==============================================================================
For methodological transparency and academic integrity, we document the
following limitations.

\begin{enumerate}
  \item \textbf{Partial reconciliation of the official universe.} The
  \code{anci-oiv-resolver} catalogue v0.5.2 covers approximately 98.7\% of the
  915 OIVs designated by Resoluci\'on Exenta N\textsuperscript{o}~87, with
  twelve state-administration entities pending reconciliation. The aggregate
  statistics are stable against the addition of twelve entities to a 915-entity
  universe; the directional conclusions do not depend on the residual.

  \item \textbf{Absence of ANCI-validated ground truth.} The findings are not
  compared against an independent reference dataset provided by ANCI. The
  Coverage-Gap metric is therefore a proxy for true operator posture, not a
  direct measurement of it. Validation against operator-internal data would
  require formal coordination with ANCI and the operators, which is outside the
  scope of independent research and is more appropriately conducted by the
  regulator itself.

  \item \textbf{Residual false-positive rate.} Post-quality-gate audit
  identified residual false-positive findings at an estimated rate of
  approximately 0.8\% of persisted findings. This sits within commonly cited
  tolerances for passive-OSINT research of the present type, but reinforces the
  design decision to keep humans in the loop for any disclosure communication.

  \item \textbf{Point-in-time snapshot.} The findings reflect the public-facing
  posture of the universe during May 2026. Operator-side remediations subsequent
  to the audit window are not reflected. The methodology is, however,
  reproducible at later dates against the same catalogue, which is the basis for
  the longitudinal future work.

  \item \textbf{Layers 1 and 2 measured; Layer 3 estimated.} Only Layers 1 and 2
  are directly measurable from passive OSINT. The Layer~3 figure is a lower-bound
  estimate derived from public evidence of disclosure-coordination behaviour. A
  full Layer~3 assessment would require operator self-attestation or a separate
  methodology.

  \item \textbf{Heterogeneous sectoral representation.} The digital-infrastructure
  and state-administration sectors together account for 62\% of the universe by
  entity count. The smaller sectors (water, transport, fuel) carry fewer entities
  and therefore lower statistical precision; conclusions about systemic patterns
  in these sectors should be treated as indicative rather than definitive.
\end{enumerate}

%==============================================================================
\section{Ethics Considerations}
\label{sec:ethics}
%==============================================================================
This research was designed and conducted in accordance with the principles of
the Menlo Report on ethical principles guiding information and communication
technology research, and the Belmont Report from which it descends. We address
each of the four Menlo principles explicitly.

\paragraph{Respect for persons}
The unit of analysis throughout is the \emph{organisational perimeter} of a
legally designated critical-infrastructure operator, not any natural person. The
study collects no personally identifiable information (PII). Where individual
human roles are mentioned in the historical-threat context
(Section~\ref{sec:related} and the introduction), they are drawn verbatim from
already-public incident reporting by third-party analysts and are not the
subject of our own data collection. No human subjects were enrolled, observed,
or contacted as research subjects.

\paragraph{Beneficence}
We weighed the benefits of measuring and publicising a systemic
disclosure-capability deficit against the risk that publication could inform an
adversary. Two design choices follow from this balance. First, the data
collection is strictly passive OSINT over information the operators themselves
made public --- DNS records, certificate-transparency logs, public Internet scan
databases, \code{/.well-known/} resources, and operators' own websites --- with
\emph{zero authentication, zero payloads, and zero exploitation}. We observe
only what an ordinary visitor could observe. Second, all operator-level detail
is held back: the paper reports aggregate, universe-scale proportions only, so
that the public artefact informs policy without functioning as a target list
(see Section~\ref{sec:disclosure}). The validation pipeline is deliberately
biased towards false negatives, accepting under-detection of real issues in
exchange for minimising the risk of misattributing a vulnerability to an
operator.

\paragraph{Justice}
The benefits and burdens of the research are equitably distributed. The
measurement applies uniformly to the entire legally defined OIV universe; no
operator or sector is singled out for disproportionate scrutiny, and no operator
is named. The principal benefit --- evidence to support a low-cost, universe-wide
regulatory intervention --- accrues to the public and to the operators
collectively, including the smaller and less-resourced entities (e.g.\ in the
health and water sectors) least able to fund independent assessment.

\paragraph{Respect for law and public interest}
The work was conducted within Chile's computer-crimes statute,
Ley~21.459~\cite{ley-21459}, under a good-faith, passive-research posture, and is
framed in accordance with the principles of
ISO/IEC~29147:2018~\cite{iso-29147}. We are transparent about methods and
limitations (Sections~\ref{sec:methodology} and~\ref{sec:limitations}) and
accountable through the named correspondence channel and the open-source release
of the catalogue tooling. We claim compliance only at the level the evidence
supports: ISO/IEC~29147 is a process standard and is not certifiable, so we
state consistency with its principles rather than operation under it.

%==============================================================================
\section{Responsible Disclosure}
\label{sec:disclosure}
%==============================================================================
Our disclosure practice is consistent with ISO/IEC~29147:2018~\cite{iso-29147}.
Operator-level findings are \emph{not} published in this paper or its companion
artefacts. Where the aggregate statistics are driven by specific operators or
infrastructure clusters, the underlying operator-level detail (identities,
specific cluster compositions, and evidence chains) is held back.

The disclosure trajectory is coordinated and staged. Where individual operators
were found to present unusually severe exposure, advisory notifications are
issued to the affected operators ahead of any public reporting. In parallel, and
consistent with the role of a national coordinator, aggregate and
sector-level findings are shared with ANCI and CSIRT-GOB as the national
coordinating body for critical-infrastructure cybersecurity --- a function
analogous to that of CISA in the United States. Operator-level detail is released
only after the coordination window has elapsed, so that operators have a
reasonable opportunity to remediate before any identifying information could
enter the public record. This paper reports aggregate findings only; the
operator-level layer is deferred to subsequent work conditioned on the
conclusion of the coordinated-disclosure process.

%==============================================================================
\section{Data and Artifact Availability}
\label{sec:artifacts}
%==============================================================================
In keeping with open-science norms for measurement research, the artefacts that
make this work reproducible are released; the artefacts that would function as a
target list for the regulated perimeter are held back pending coordinated
disclosure (a \emph{partial-availability} declaration in the style adopted by
the Internet-measurement community).

\paragraph{Released}
The catalogue tool \code{anci-oiv-resolver} v0.5.2 is published under the Apache
License 2.0. It is available on npm as the package
\texttt{anci-oiv-\cb resolver@0.5.2}
--- distributed with an SLSA provenance attestation generated via an OIDC trusted
publisher --- with source at
\texttt{github.com/\cb raceksd-source/\cb anci-oiv-resolver}
and an archived release deposit on Zenodo (DOI:
\texttt{10.5281/\cb zenodo.20501614}). The tool provides the canonical RUT-to-domain
mapping covering approximately 98.7\% of the OIV universe and is the foundation
on which the Layer~1 census was conducted. The Layer~1 \code{security.txt} census
(RFC~9116~\cite{rfc-9116}) is provided as a companion dataset, enabling
independent reproduction of the headline 1.7\% figure.

\paragraph{Held back}
Operator-level findings --- specific operator identities, cluster compositions,
the per-operator Layer~2/3 assessments, and the detailed evidence chains behind
the aggregate statistics --- are held back pending the conclusion of the
coordinated-disclosure process described in Section~\ref{sec:disclosure}. This
asymmetry is deliberate: the released artefacts are sufficient to reproduce the
\emph{population-level} claims of this paper, while the withheld artefacts are
those whose premature release would disproportionately benefit an adversary
relative to a defender.

The intended consequences of the release are threefold. First, independent
researchers can reproduce the universe enumeration that anchors the Phase~1
audit without re-deriving the catalogue from the legal designation in
Resoluci\'on Exenta N\textsuperscript{o}~87. Second, researchers operating under
alternate methodologies --- active reconnaissance with operator authorisation,
internal-audit access, or sectoral self-reporting --- can apply their methods to
the same legally defined universe, producing cross-method evidence that this
paper cannot produce on its own. Third, the catalogue is positioned as a
community artefact open to correction through its public issue tracker. The
intent is that the Chilean critical-infrastructure perimeter ceases to be a
research bottleneck and becomes, instead, a research commons.

%==============================================================================
\section{Conclusion}
\label{sec:conclusion}
%==============================================================================
Chile now has, on paper, a critical-infrastructure cybersecurity framework: a
framework law (Ley~21.663), a regulator (ANCI), a legally designated regulated
perimeter (915 OIVs), and a first cohort of binding obligations falling due in
mid-2026. What it does not yet have is the operational substrate that makes such
a framework effective. A full-universe, passive-OSINT census shows that only
1.7\% of the regulated perimeter publishes a verifiable disclosure channel, that
84\% exhibit email-authentication misconfiguration, and that a non-trivial
fraction advertise end-of-life or vulnerable software. Benchmarked against five
jurisdictions that each closed a comparable gap within two years of a binding
mandate, Chile sits roughly eight years behind the leading cohort --- but with
the corresponding opportunity to compress that iteration into a single adoption
step. The remedy is well established and low cost: a DMARC mandate analogous to
CISA BOD 18-01, a published disclosure-contact registry, a regulator-operated
report-analyser service, and an annual comply-or-explain Coverage-Gap report.
The Coverage Gap is not a measure of how secure Chilean critical infrastructure
is; it is a measure of whether the next discovered vulnerability has anywhere to
go. Today, for 98.3\% of the regulated perimeter, it does not.

\subsection{Future work}
\label{sec:futurework}
This paper is the opening contribution of a sustained research programme.
Committed future tracks include: a sectoral deep-dive expanding the
Shodan-enriched subset to a target of at least one hundred entities --- which,
under typical sampling distributions, narrows the 95\% confidence interval on
the Level~C stack-age claim to within approximately $\pm5$ percentage points ---
and prioritising the sectors flagged here (banking and finance; health; and the
digital-infrastructure cohort); a systemic-risk and infrastructure-dependency
analysis characterising shared-upstream-dependency patterns across the universe,
released on a timeline consistent with the coordinated-disclosure window and
with operator-level specifics held back until that process concludes; a
dedicated methodological treatment of the passive-OSINT-plus-multi-layer-validation
pipeline with a reproducible evaluation harness; and a longitudinal re-audit of
the same 915-entity universe at a defined cadence beginning Q3 2026, providing
the empirical baseline against which the impact of any regulatory intervention
can be quantified. Researchers, regulators, journalists, civil-society
organisations, and operator-side teams who wish to extend, contradict, or
replicate these findings are invited to do so; correspondence is welcomed at the
address in the author block.

%==============================================================================
% Bibliography
%==============================================================================
\balance
\bibliographystyle{IEEEtran}
\bibliography{refs}

\end{document}